\documentclass[twocolumn,aps,prb,longbibliography,superscriptaddress]{revtex4-2}
\usepackage{graphicx}
\usepackage{dcolumn}
\usepackage{bm}
\usepackage{amssymb, amsmath}
\usepackage{color}
\usepackage{slashed} 
\usepackage{nicefrac}
\setcitestyle{super}
\begin{document}
	\title{Optical and transport properties of NbN thin films revisited}

	\author{S. Kern}
	\affiliation{Department of Experimental
		Physics, Comenius University, SK-84248 Bratislava, Slovakia}
	\author{P. Neilinger}
	\affiliation{Department of Experimental
		Physics, Comenius University, SK-84248 Bratislava, Slovakia}
	\affiliation{Institute of Physics, Slovak Academy of Sciences,
		D\'{u}bravsk\'{a} cesta, SK-84511 Bratislava, Slovakia}
	\author{M. Pol\'{a}\v{c}kov\'{a}}
	\affiliation{Department of Experimental
		Physics, Comenius University, SK-84248 Bratislava, Slovakia}
	\author{M. Bar\'{a}nek}
	\affiliation{Department of Experimental
		Physics, Comenius University, SK-84248 Bratislava, Slovakia}
	\author{T. Plecenik}
	\affiliation{Department of Experimental
		Physics, Comenius University, SK-84248 Bratislava, Slovakia}
	\author{T. Roch}
	\affiliation{Department of Experimental
		Physics, Comenius University, SK-84248 Bratislava, Slovakia}
	\author{M.~Grajcar}
	\affiliation{Department of Experimental
		Physics, Comenius University, SK-84248 Bratislava, Slovakia}
	\affiliation{Institute of Physics, Slovak Academy of Sciences,
		D\'{u}bravsk\'{a} cesta, SK-84511 Bratislava, Slovakia}

\begin{abstract}
Highly disordered NbN thin films exhibit promising superconducting and optical properties.
Despite extensive study, discrepancies in its basic electronic properties persist.
Analysis of the optical conductivity of disordered ultra-thin NbN films, obtained from spectroscopic
ellipsometry by standard Drude-Lorentz model, provides inconsistent parameters. We argue that this
discrepancy arises from neglecting the presence of quantum corrections to conductivity in the IR range.
To resolve this matter, we suggest a modification to the Drude-Lorentz model, incorporating quantum
corrections. The parameters obtained from the modified model are consistent with transport and superconducting measurements. The revisited
values describing conduction electrons, which differ significantly from commonly adopted ones, are the
electron relaxation rate $\Gamma\approx1.8~\textrm{eV}/\hbar$, the Fermi velocity
$v_F \approx 0.7 \times 10^{6}~\textrm{ms}^{-1}$ and the electron density of states $N(E_F)=2~$states of
both spins/eV/$V_{\textrm{f.u.}}$.

\end{abstract}
\maketitle

\section{Introduction}
\label{ch:intro}

Niobium nitride (NbN) stands out for its excellent properties, including chemical stability,
hardness, optical and superconducting characteristics.\cite{toth2014transition, pogrebnjak2016structure,
banerjee2018optical, hazra2016superconducting} Its relatively high superconducting critical
temperature and large sheet resistance make it suitable for applications such as superconducting nanowire single
photon detectors (SNSPDs) \cite{gol2001picosecond} and kinetic inductance travelling wave parametric amplifiers.
\cite{adamyan2016superconducting} Transition metal nitrides, including NbN, nowadays garner interest as
plasmonic materials,\cite{karl2020optical} exhibiting double epsilon-near-zero (ENZ) behaviour.
\cite{bower2021tunable} This behaviour means that the real part of their dielectric function,
$\epsilon_r(\omega)$, becomes zero at two frequencies below the UV range. ENZ materials enable strong
interaction of light with plasma oscillations, offering a wide range of possibilities
in photonics.\cite{wu2021epsilon} NbN has drawn attention as an ENZ material due to its tunable plasma
frequency through composition adjustments.\cite{ran2021stoichiometry} Moreover, the optical response
directly influences the efficiency of SNSPDs,
emphasizing the importance of optical characterization of NbN thin films, as highlighted in Ref.~\citenum{semenov2009optical}.

Despite more than 50 years of extensive study of NbN films, significant disagreement persists with regards to
some of their fundamental properties. The primary source of mismatch arises from seemingly
contradictory results obtained through different measurement pathways.\cite{semenov2009optical,sidorova2020electron}
First, the characterisation of the disorder in NbN is routinely obtained from transport and Hall effect measurements. This is done via the Ioffe-Regel parameter $k_F l$, where $k_F$ is the Fermi wavevector and $l$ is the electron mean free path. It is well known that $k_Fl$ close to unity can be obtained in thin NbN films, i.e. $l$ is comparable to electron wavelength, and such highly disordered films are approaching metal-insulator transition (MIT).\cite{chockalingam2008superconducting,pellan1990temperature, chand2012phase} Alternatively, this criterion can be expressed via electron scattering rate in energy units $\hbar\Gamma$, which is comparable to the Fermi energy. Therefore, in highly disordered metals, $\hbar\Gamma$ is expected to be a couple eV. Second, $\hbar\Gamma$ can be obtained from optical response as well, namely as a parameter of the Drude model for the dielectric function of conducting electrons. However, these measurements suggest $\hbar\Gamma\approx0.33~$ as obtained from fitting a Drude-Lorentz model to ellipsometric data in the visible range,\cite{semenov2009optical,kuz1983determination} which is an order of magnitude smaller than expected.

Thin NbN films, especially its $\delta-$phase, have superior superconducting properties,\cite{chand2012transport} and can be deposited by various methods,\cite{kafizas2013cvd,volkov2019superconducting,
ziegler2012superconducting}. They typically exhibit a polycrystalline structure with grains of various sizes,
each possessing a relatively well-defined cubic lattice interrupted by vacancy defects.
\cite{lin2013characterization, toth2014transition} The grain boundaries consist of disordered NbN
alloy, often containing oxygen, as well.\cite{lengauer1986herstellung,cabanel1990correlations}
Despite their granularity, NbN thin films can be considered as
homogeneously disordered metal, especially regarding their optical response. This is natural in the case of high intergrain conductivity\cite{beloborodov2007granular}
or for mean free path that is small in comparison to the grain size. \cite{reiss1986grain,pellan1990temperature}

It is known that the presence of disorder in metals, either granular or homogenous, leads to quantum corrections (QCs)
to the Drude conductivity. In highly disordered metals, the density of states (DOS) at the Fermi level is suppressed, suppressing their conductivity, as well.
\cite{Altshuler1985} The correction to the real part of conductivity in
3D homogeneously disordered films, as obtained by Altshuler and Aronov in Ref.~\citenum{Altshuler1985}
as well as from scaling arguments,\cite{mcmillan1981scaling} can be expressed in the following unified form
\cite{neilinger2019observation, kaveh1982universal}
\begin{equation}
\delta \sigma(\Omega) = -\mathcal{Q}^2\sigma_0\left(1-\sqrt{\frac{\Omega}{\Gamma}}\right).
\label{eq:AA}
\end{equation}
Here, $\sigma_0$  is the Drude conductivity and  $\mathcal{Q}$ is the strength of the quantum corrections, also referred to as quantumness.
\cite{neilinger2019observation} The electron relaxation rate is defined as reciprocal relaxation time $\Gamma=1/\tau$. The quantity $\Omega$ is determined
by incident photon energy $\hbar\omega$, and
temperature $\sim k_BT$, and can be approximated as \cite{Altshuler1985,neilinger2019observation}
\begin{equation}
\Omega = \sqrt{\omega^2+(\pi k_BT/\hbar)^2}
\label{eq:O}
\end{equation}
which effectively leads to the temperature smearing of the square root singularity in Eq. (\ref{eq:AA}) at low photon energies.
Although this behaviour is routinely observed at energies
of the order of meV, it is rarely taken into account in optical response analysis.
\cite{lee1995reflectance} Notably, in the study by Neilinger et al.,\cite{neilinger2019observation} the square root corrections
(\ref{eq:AA}) were observed up to optical frequencies in MoC thin films. Although not directly measured,
a numerical study proposed a similar square root behavior in NbN.\cite{kern2021numerical} 

In the following, we argue that these corrections dramatically alter the optical properties of NbN films and explain the
ENZ phenomenon. Modeling the optical conductivity by a quantum-corrected Drude-Lorentz model, we can determine various
quantities, including the diffusion coefficient and the superconducting coherence length in agreement with the transport
measurements. Moreover, the determined carrier density agrees with the DFT results in the literature. Additionally, we compare this model to commonly used models of
the NbN dielectric function. The experimental data used to establish a consistent set of parameters, were obtained from
spectroscopic ellipsometry (SE) on $\delta-$NbN films of various thicknesses. Independent confirmation was provided by transport and magneto-resistance measurements.

\section{Optical properties}
\label{ch:optic}

\begin{figure}
	\centering
	\includegraphics[width=8.7cm]{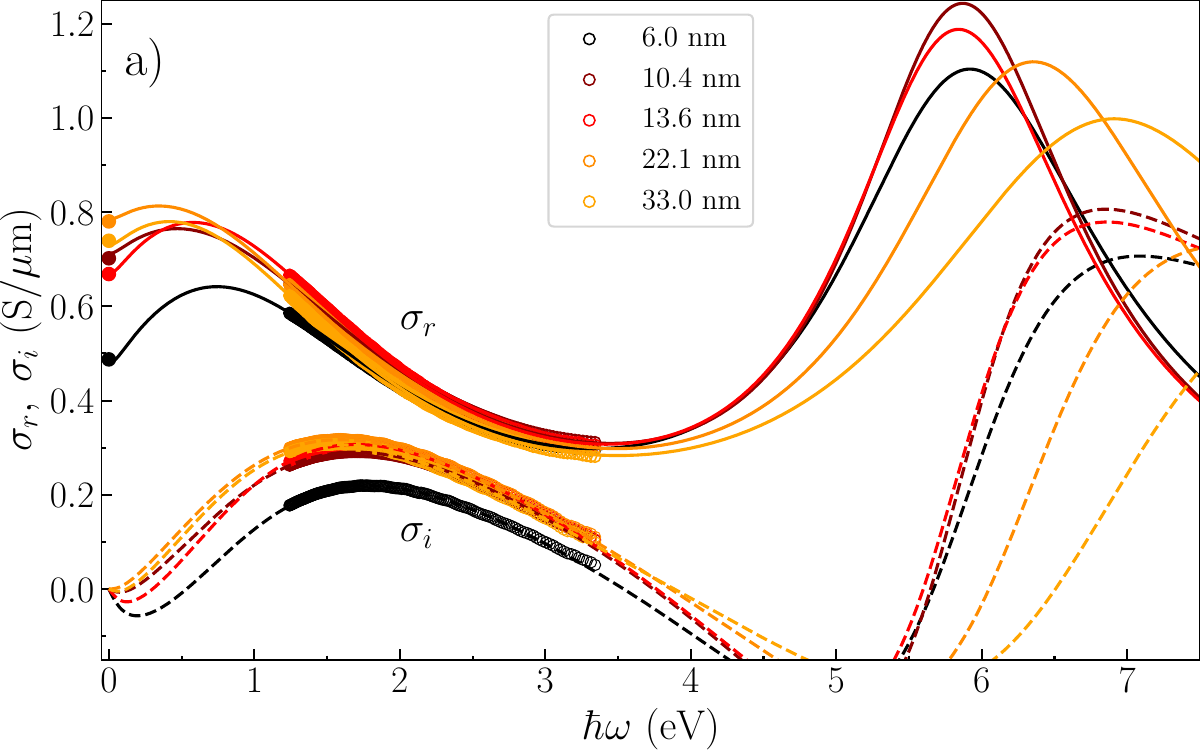}
	\includegraphics[width=8.8cm]{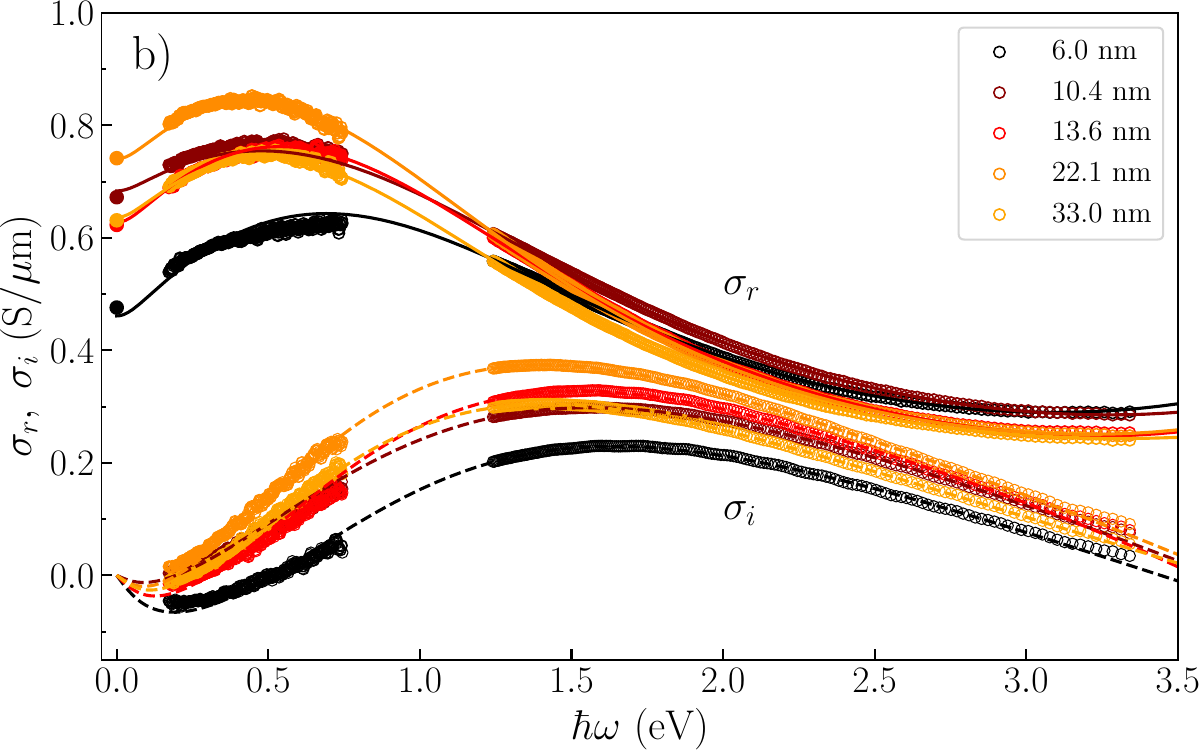}

	\caption{Open circles: Real and imaginary parts of the optical conductivity for NbN films of various thicknesses, determined by spectroscopic ellipsometry. Thin lines: Fits to Eq.~(\ref{eq:TheModel}). Solid circles: Room-temperature DC conductivities measured using the van der Pauw method.
			a) Optical conductivity from visible-range SE measured shortly after deposition.
			b) Visible and mid-IR range SE data measured one year after deposition. The visible-range data alone are used in the fit.}
	\label{fig:se}
	
\end{figure}
The optical properties of NbN thin films were investigated utilizing SE in the visible range from 300 to 800 nm
and subsequently mid-infrared (mid-IR) range from 1.7 to 7.2 $\mu\textrm{m}$. Typically, SE is employed to determine the parameters
(e.g., layer thickness, plasma frequency, relaxation time etc.) of a chosen optical model for
the studied sample. In this study, the optical constants were evaluated
directly from the SE data at each wavelength separately by a model-independent procedure
(for details see Appendix \ref{app:eoc} or Ref.~\citenum{neilinger2021study}).

The visible range SE measurements were carried out shortly after the deposition of the films. The determined optical conductivities are shown as circles in Fig.~\ref{fig:se} a).
The mid-IR SE measurements shown in  Fig.~\ref{fig:se} b)  were conducted with the newly installed Infrared Spectroscopic Ellipsometer (SENDIRA), a year after the original visible-range measurements.
During this time, the samples underwent slight degradation, as evidenced by the increased sheet resistance  (see Appendix \ref{app:mir}). Therefore, we repeated the visible range measurements (Fig. \ref{fig:se} b)). This ensured that the mid-IR data and the new visible-range data seamlessly aligned at the boundaries of their spectral ranges. The solid circles represent DC conductivities ($\sigma_{DC}$) determined from the sheet resistance measured using the van der Pauw method.
The transport properties studied in the subsequent chapter, measured immediately after the deposition, are fully consistent only with the analysis based on the initial visible-range data. However, the mid-IR measurements presented in this chapter provide valuable support to the analysis. Most importantly, the mid-IR data reflect the key features of the proposed quantum-corrected Drude model.

The suppression of $\sigma_r(\omega)$ revealed by mid-IR SE measurements results in a peak with maxima at $0.5-1$ eV. This 'anti-Drude' behavior, also known as an anomalous or displaced Drude peak,
was also predicted by the numerical study in Ref.~\citenum{fratini2021displaced} and extrapolation based on the KK relations.\cite{kern2021numerical} Thus, we attribute  the decrease in the real part of the optical conductivity in the visible spectral range to the Drude-like $1/\omega^2$ behaviour
typical for metals with finite relaxation rate $\Gamma$. As a consequence, $\Gamma$ in the studied NbN films is 1.5-2 eV.

In Ref.~\citenum{bower2021tunable}, the peak in the optical conductivity of 
NbN films was explained in terms of an effective medium emerging from granular NbN dissolved in an insulating
NbO matrix. We discuss this approach in Appendix~\ref{app:emt}, where we argue that it is not appropriate for NbN, as it leads to unphysical conclusions.\cite{cocker2017microscopic} In Ref.~\citenum{cocker2017microscopic}, the peak was obtained within the Drude-Smith model,
which we analyze in Appendix~\ref{app:DSM}.

In Ref.~\citenum{lee1995reflectance}, the $\sigma_r(\omega)$ of metallic films close to MIT exhibited a square root dependence at frequencies below IR range, which was ascribed to quantum corrections to the Drude conductivity due to disorder. The imaginary part $\sigma_i(\omega)$ corresponding to square root in the $\sigma_r(\omega)$ should reach negative values at low energies (THz and IR range). The mid-IR ellipsometry carried out on our NbN films (Fig. \ref{fig:se} b)) provide $\sigma_i(\omega)$, which indeed changes sign for all samples expect the 22 nm thick sample. In THz range, the negative $\sigma_i(\omega)$ was reported in Ref.~\citenum{pracht2013electrodynamics}, and its association with disorder was pointed out in Ref.~\citenum{cheng2016anomalous}. In order to describe the $\sigma_r(\omega)$ in the optical range, a "localization-modified Drude model" was suggested as a simple multiplication of the Drude formula with a
square root term similar to Eq.~(\ref{eq:AA}).\cite{kaveh1982universal} However, this model leads to incorrect behaviour of the conductivity
at high frequencies, which should drop as $\omega^{-2}$. Moreover, this model
poorly fits our data. The same square root behaviour of $\sigma_r(\omega)$ up to the visible range was observed in Ref.~\citenum{neilinger2019observation}. But, it was suggested here that the influence of disorder should
disappear at frequencies of the order of the relaxation rate. A square root correction was
smoothly joined to the bare Drude conductivity at a crossover frequency, and an excellent agreement with the SE data was
further confirmed by independent transmission measurements, suggesting that the presence of the QCs must be considered in an optical model.\cite{neilinger2019observation} Therefore, we apply the quantum-corrected Drude conductivity
\begin{equation}
\begin{aligned}
\sigma_{QCD}(\omega) =\frac{\sigma_0}{1+(\frac{\omega}{\Gamma})^2}\left(1-\mathcal{Q}^2\left(1-\sqrt{\frac{\Omega}{\Gamma}}\right)e^{-2(\nicefrac{\omega}{\Gamma})^2}\right),
\end{aligned}
\label{eq:QCD}
\end{equation}
where quantum correction exponentially vanishes at a scale $\Gamma/2$. This scale was fixed after its release during data fitting consistently yielded a uniform result. This may not be the case for much more disordered samples.\cite{neilinger2021study}  The value of $\Omega$ is given by Eq. (\ref{eq:O}) and  $T=300~$K, as the spectra were measured at room temperature.

In Ref.~\citenum{kern2021numerical}, a similar NbN film was studied, and a conductivity peak in the UV range was
determined by means of numerical extrapolation. This spectral weight was attributed to inter-band transition at
$\hbar\omega\approx 5-7~$eV. The presence of the inter-band transition, was observed through optical measurements \cite{sanjines2006electronic,ran2021stoichiometry} and various ab-initio simulation, as well.\cite{fong1972pseudopotential,pfluger1985dielectric,ran2021stoichiometry}
We model the 5-7 eV inter-band transition by a Lorentzian peak centered at $\omega_1$, with strength $\sigma_1$ and width $\Gamma_1$. 
Finally, we obtained the following model for the real part of the complex conductivity $\sigma_r(\omega)+i\sigma_i(\omega)$
\begin{equation}
\begin{aligned}
\sigma_r(\omega) = \sigma_{QCD}(\omega)+\frac{\sigma_1}{1+\left(\frac{\omega_1^2-\omega^2}{\omega\Gamma_1}\right)^2},
\end{aligned}
\label{eq:TheModel}
\end{equation}
\begin{equation}
\begin{aligned}
\sigma_i(\omega) = &\mathcal{H}[\sigma_{\textrm{QCD}}(\omega)]- \frac{\omega_1^2-\omega^2}{\omega\Gamma_1} \frac{\sigma_1}{1+\left(\frac{\omega_1^2-\omega^2}{\omega\Gamma_1}\right)^2}- \\ &(\epsilon_\infty-1)\epsilon_0\omega.
\end{aligned}
\label{eq:im}
\end{equation}
Here,
$\mathcal{H}[\sigma_{QCD}(\omega)]$ denotes the Hilbert transform of the quantum-corrected Drude model, i.e. the transform of the first term of $\sigma_r(\omega)$ in Eq. (\ref{eq:TheModel}), manifesting the Kramers-Kronig relations. The transform can be performed numerically, or by approximative analytical formula, which we derived in Appendix~\ref{app:app}. The second term of $\sigma_i$ is the imaginary part corresponding to the inter-band transition.
This model of $\sigma_r(\omega)$ does not consider the inter-band transitions of bound electrons at high energies ($> 7~$eV). Their effect on the imaginary conductivity is included in the third term in Eq.~(\ref{eq:im}) via the parameter $\epsilon_\infty=1.6$ estimated in Appendix~\ref{app:elec}.

The proposed model (thin lines in the Figs.~\ref{fig:se} a) and b)) produces an excellent
fit to both the real and the imaginary parts of the conductivity. Furthermore, utilizing only
the visible range data the extrapolation by the fit to the proposed model is closely matched by the mid-IR optical conductivity  (Fig. \ref{fig:se} b)), strongly supporting the model. The static conductivities were perfectly recovered by the fit, as well. The parameters of the best fit for the original and repeated analysis are listed in Table~\ref{tab:params} and Table~\ref{tab:params2}, respectively. The Drude conductivity $\sigma_0$ and the parameters of the inter-band transition peak exhibit variance, making it difficult to clearly observe their dependence on thickness. However, despite the variability in the scattering rate $\Gamma$, a subtle increasing trend with decreasing thickness can be inferred.  A more pronounced thickness dependence is expected for thinner samples ($<6~$nm), as demonstrated, for instance, in Ref. \citenum{zhang2022superconductivity}. As $\Gamma$ rises, 
$\mathcal{Q}$ increases, too. This is consistent with expression for quantumness $\mathcal{Q}\approx 1/k_Fl$,
where $l=v_F/\Gamma$ and $v_F=\hbar k_F/m_e$ is
the Fermi velocity, which we estimated as $v_F=\sqrt{\hbar\Gamma/(\mathcal{Q}m_e)}\approx 0.7-0.8\times10^6~$ms$^{-1}$.\cite{Altshuler1985,kaveh1982universal}

The presence of the peak at 5-7 eV agrees with the DFT calculation presented in Ref. \citenum{pfluger1985dielectric}, where a peak at $\approx 7~$eV was found in the imaginary part of dielectric function $\epsilon_i$, determined from the joint density of states. The reported value in the peak $\epsilon_i\approx5$
corresponds to $\sigma_r=\omega\epsilon_0\epsilon_i\approx0.5~\textrm{S}/\mu\textrm{m}$, which is in good agreement to the weight of the Lorentzian found for the thickest sample $\sigma_1=0.68~\textrm{S}/\mu\textrm{m}$.
Similarly, the calculation of the joint density of states predicted transitions between the three highest occupied
bands, leading to a peak in the dielectric function at $\approx 1~$eV. Therefore, it is tempting to
assign the anomalous Drude peak to the inter-band transitions, as was done in Refs.~\citenum{semenov2009optical, kuz1983determination}. However, this approach results in inconsistencies of conduction electron contribution, such as $\hbar\Gamma\sim0.1~$eV. The fact that $\hbar\Gamma$ is approximately 10 times larger is particularly evident in the angle-resolved photoemission spectroscopy ARPES measurement, where
the disorder and/or thickness-enhanced scattering smears the electron band-structure at the scale
$\hbar\Gamma \approx 1.5~$eV.\cite{yu2021momentum}
Furthermore, in Ref. \citenum{babu2019electron} based on DFT simulations, it is directly stated, that ``\textit{in $\delta$-NbN,
all the calculated phonon frequencies are positive only when
an abnormally large value of the electronic band smearing
width ($\ge$0.15 Ry) is used.}" meaning that such smearing is
necessary to stabilize the crystalline structure of $\delta-$NbN. As a consequence of the large value of $\Gamma$,
the individual electronic bands, which are separated by
energy less than $\hbar\Gamma$, lose their identity and merge together. Therefore, these inter-band transitions should not be present in the spectra.
Also, modelling the displaced Drude peak as an inter-band transition leads to a puzzling shift of its central frequency to
higher energies with decreasing thickness, whereas for our model it is explained via the increase of the
quantumness $\mathcal{Q}$.
\begin{figure}
	\centering
	\includegraphics[width=8.6cm]{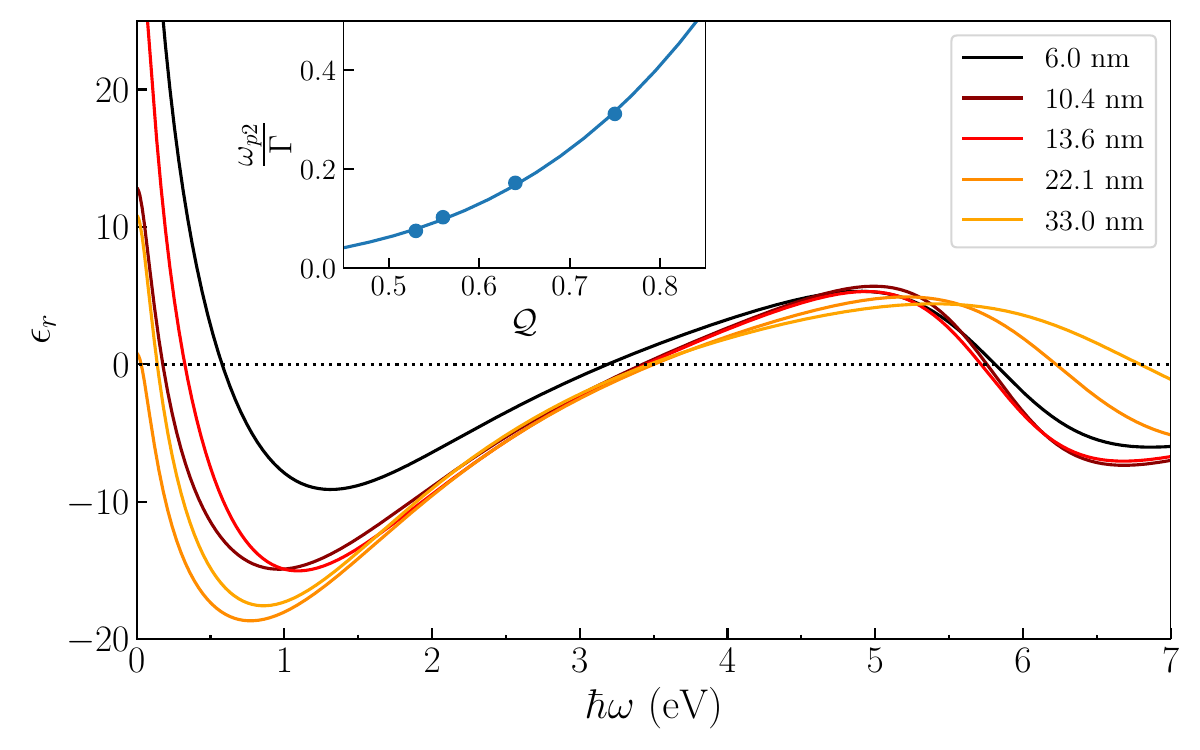}
	\caption{The real part of the dielectric function $\epsilon(\omega)$ corresponding to the conductivities in Fig.~\ref{fig:se} a). The inset shows the lower plasma frequencies (frequencies at which $\epsilon(\omega)=0$) dependent on quantumness $\mathcal{Q}$. The solid line is a plot according to Eq.~(\ref{eq:wp2})}
	\label{fig:eps}
\end{figure}

The corresponding real part of the dielectric function (Fig.~\ref{fig:eps}) given as $\epsilon_r(\omega) = 1 - \sigma_i(\omega)/\epsilon_0\omega$,
exhibits the discussed double ENZ feature. For $\mathcal{Q} = 0$,
the model gives (for $\epsilon_r(\omega)$ and for the ordinary screened plasma
frequency $\omega_p$) the well-known results
\begin{equation}
\begin{aligned}
\epsilon_r(\omega) \approx \epsilon^\prime_\infty-&\frac{\sigma_0/\epsilon_0\Gamma}{(1+(\omega/\Gamma)^2)},\\
\omega_p^2 = &\frac{\sigma_0\Gamma}{\epsilon_0\epsilon_\infty^\prime},
\end{aligned}
\label{eq:eps}
\end{equation}
where $\epsilon_\infty^\prime = \epsilon_\infty + \sigma_1\Gamma_1/(\epsilon_0\Omega_1^2)$ contains bound electrons contribution $\epsilon_\infty$ evaluated in Appendix~\ref{app:elec} and contribution from the inter-band transition at 5-7 eV.
From Eq.~(\ref{eq:imx}), it can be shown, that for low frequencies,
the $\sigma_i(\omega)$ is dominated by the square root term. This modifies $\epsilon_r(\omega) $ and leads to the newly unveiled second plasma frequency $\omega_{p2}$
\begin{equation}
\begin{aligned}
\epsilon_r(\omega) \approx \epsilon^\prime_\infty-&\frac{\sigma_0}{\epsilon_0\Gamma}\left(1-\mathcal{Q}^2\sqrt{\frac{\Gamma}{\omega} }      \right), \\
 \omega_{p2} &\propto\Gamma \mathcal{Q}^4.
\end{aligned}
\label{eq:wp2}
\end{equation}
The second, lower plasma frequency $\omega_{p2}$ increases as $\mathcal{Q}^4$, whereas the regular plasma frequency $\omega_p$ slightly decreases with
$\mathcal{Q}$. In the inset of Fig.~\ref{fig:eps}, the second plasma frequencies of
the samples are plotted. The thickest sample with the lowest $\mathcal{Q}$ does not reach zero, as the
temperature smearing sufficiently suppresses the square root behaviour, which is not taken into account in Eq.~(\ref{eq:wp2}).

\begin{table}[h!]
	\centering
	\renewcommand{\arraystretch}{1.5}
	\begin{tabular}{>{\centering\arraybackslash}p{0.8cm} |>{\centering\arraybackslash}p{1.5cm} >{\centering\arraybackslash}p{1.0cm} >{\centering\arraybackslash}p{0.7cm} >{\centering\arraybackslash}p{1.5cm} >{\centering\arraybackslash}p{1.1cm}
	>{\centering\arraybackslash}p{1.0cm} }
		\hline
		$d$& $\sigma_0  $ & $\hbar\Gamma$ &  $\mathcal{Q}$& $\sigma_1 $ & $\hbar\Gamma_1$& $\hbar\omega_1$\\ 
		
		$(\textrm{nm})$& $(S\mu\textrm{m}^{-1})$ & $(\textrm{eV})$ &  $(1)$& $(S\mu\textrm{m}^{-1})$ & $(\textrm{eV})$& $(\textrm{eV})$\\ 
		\hline\hline
		6.0  & 0.88 & 1.86 & 0.75 & 0.72 & 2.38 & 5.64\\ 
		10.4 & 0.94 & 1.89 & 0.56 & 0.77 & 2.04 & 5.51\\
		13.6 & 1.01 & 1.80 & 0.64 & 0.81 & 1.80 & 5.36\\
		22.1 & 0.95 & 1.84 & 0.45 & 0.77 & 2.23 & 5.74\\
		33.0 & 0.95 & 1.73 & 0.53 & 0.68 & 3.10 & 6.20\\
		\hline
	\end{tabular}
	\caption{Paremeters of optical model (\ref{eq:TheModel}) and (\ref{eq:im}) providing the best fit to the experimental data.}
	\label{tab:params}
\end{table}

\section{Transport and superconducting properties}
\label{ch:tas}

\begin{table}
	\centering
	\renewcommand{\arraystretch}{1.5}
	\begin{tabular}{>{\centering\arraybackslash}p{0.9cm} |>{\centering\arraybackslash}p{1.0cm} >{\centering\arraybackslash}p{1.3cm} >{\centering\arraybackslash}p{0.9cm} >{\centering\arraybackslash}p{1.1cm} >{\centering\arraybackslash}p{1.1cm}>{\centering\arraybackslash}p{1.6cm}
			}
		\hline 
		$d$& $L$ & $\rho$ &  $R_\square$ & $RRR$  & $T_c $&$ 2k_BT_c$\\ 
		
		$(\textrm{nm})$& $(\textrm{nm})$ & $(\textrm{gcm}^{-3})$ &  $(\Omega)$ & $(1)$  & $(K)$&$(\textrm{meV})$\\
		
		\hline\hline
		6.0  & 9  & 7.8 & 340 & 0.76 & 11.68 & 2.0\\ 
		10.4 & 12 & 7.8 & 136 & 0.88 & 13.15 & 2.3\\
		13.6 & 12 & 7.8 & 107 & 0.81 & 13.64 & 2.3\\
		22.1 & 11 & 7.9 & 58  & 0.89 & 14.26 & 2.5\\
		33.0 & 12 & 7.8 & 41  & 0.83 & 13.71 & 2.4\\
		\hline
	\end{tabular}
	\caption{Properties of thin NbN films obtained from X-ray measurements and temperature-dependent transport measurements. RRR is obtained as $R_\square/R_\square(20~\textrm{K})$, where $R_\square$ is the room temperature sheet resistance.}
	\label{tab:meas}
\end{table}

\begin{table*}[t]
	\centering
	\renewcommand{\arraystretch}{1.5}
	\begin{tabular}{>{\centering\arraybackslash}p{0.65cm} >{\centering\arraybackslash}p{0.65cm} >{\centering\arraybackslash}p{1.2cm} >{\centering\arraybackslash}p{1.9cm} >{\centering\arraybackslash}p{1.45cm} >{\centering\arraybackslash}p{1.25cm} >{\centering\arraybackslash}p{1.15cm} >{\centering\arraybackslash}p{1.25cm} >{\centering\arraybackslash}p{1.65cm} >{\centering\arraybackslash}p{1.55cm} >{\centering\arraybackslash}p{1.55cm} >{\centering\arraybackslash}p{1.4cm}}
		\hline
		$d$  	            & $k_Fl$& $D_{opt}$ & $D_{B_{c2}}$  &$l$ & $v_F$&$\xi_{0}$ & $\xi_{GL}$ & $B_{c2}$& $N(E_F)$&$E_F-E_c$ &$n$\\
		
		$ (\textrm{nm})$    & (1)   & $(\textrm{cm}^2\textrm{s}^{-1})$ & $(\textrm{cm}^2\textrm{s}^{-1})$& $(\textrm{\AA})$ & $(10^6 \textrm{ms}^{-1})$&$(\textrm{nm})$&$(\textrm{nm})$&$(\textrm{T})$&$(\textrm{eV}^{-1}V_{\textrm{f.u.}}^{-1})$&$(\textrm{eV})$&$(10^{28}\textrm{m}^{-3})$\\
		
		& $1/\mathcal{Q}$& $\frac{\hbar}{3\mathcal{Q}m_e}$ & $-\frac{4k_{B}}{\pi e}\left(\frac{ \partial B_{c2}}{\partial T}\right)^{-1}_{T_c}$  & $\sqrt{\frac{\hbar}{\mathcal{Q}\Gamma m_e}}$ & $\sqrt{\frac{\hbar\Gamma}{\mathcal{Q}m_e}}$ & $\frac{\hbar v_F}{\pi\Delta}$& $0.85\sqrt{\xi_{0}l}$&$\Phi_0/2\pi\xi_{GL}^2$& $\frac{3\sigma_0\mathcal{Q}m_e}{\hbar e^2}$&$\frac{\hbar\Gamma}{2\mathcal{Q}}$ &$\frac{\sigma_0 \Gamma m_e}{e^2}$\\
		
		\hline\hline
		6.0     & 1.33  & 0.51   & 0.73 &	2.34  & 0.66   & 69.2  & 3.44 &	27.0 & 2.27 & 1.24	& 8.82\\
		10.4     & 1.78  & 0.69   & 0.68 &	2.68  & 0.77   & 70.2  & 3.71 &	23.1 & 1.81 & 1.69	& 9.57\\
		13.6     & 1.56  & 0.60   & 0.60 &	2.57  & 0.70   & 64.1  & 3.47 &	26.5 & 2.23 & 1.41	& 9.80\\
		22.1     & 2.22  & 0.86   & 0.63 &	3.03  & 0.85   & 71.1  & 3.97 &	20.2 & 1.47 & 2.04	& 9.42\\
		33.0     & 1.88  & 0.73   & 0.57 &	2.88  & 0.76   & 66.1  & 3.73 &	22.9 & 1.73 & 1.63	& 8.86\\
		\hline
	\end{tabular}
	\caption{Electronic properties calculated from the fit parameters of the proposed optical
		model to optical conductivity. For comparison, besides the calculated diffusivity $D_{opt}$, diffusivity $D_{B_c2}$ obtained from the temperature dependence of
		critical field $B_{c2}(T)$ is listed, too. For the BCS coherence length $\xi_0$ we used the estimate of superconducting
		gap from Table~\ref{tab:meas}: $\Delta\approx2k_BT_c$.}
	\label{tab:prop}
\end{table*}

The introduced optical model, fitted to ellipsometric data, predicts the DC conductivity that was
independently evaluated as $\sigma_{DC}(T=300~\textrm{K})= 1/(R_\square d)$. Here, $R_\square$ is
the sheet resistance measured by the van der Pauw method at room temperature, and $d$ is the thickness
of the sample determined by the X-Ray reflection (XRR) measurements. Both values are listed in table \ref{tab:meas}.
The resulting DC conductivity is plotted in Fig.~\ref{fig:se} as dots, and they are perfectly
recovered by the low-frequency part of the optical conductivity fit. Here, we emphasize, that the
DC conductivity was not utilized during the fitting procedure.

In literature, it is common to compute the electronic parameters (i.e. $k_Fl$) from the measured DC conductivity
$\sigma_{DC}= 1/(R_\square d)$, assuming it is equal to the Drude conductivity $\sigma_0$. However,
within our approach, the measured DC conductivity is $\sigma_{DC} = \sigma_0(1-\mathcal{Q}^2)$, therefore it can not be interchanged with the Drude conductivity $\sigma_0$.
Following Refs.~\citenum{Altshuler1985,lee1995reflectance}, we equate the quantumness $\mathcal{Q}$ to $1/k_Fl$,
which enables us to easily estimate the Ioffe-Regel parameter as well as the diffusivity
$D = v_Fl/3=\hbar k_F l/(3m_e)=\hbar/(3\mathcal{Q}m_e)$.

\begin{figure}
	\centering
	\includegraphics[width=8.6cm]{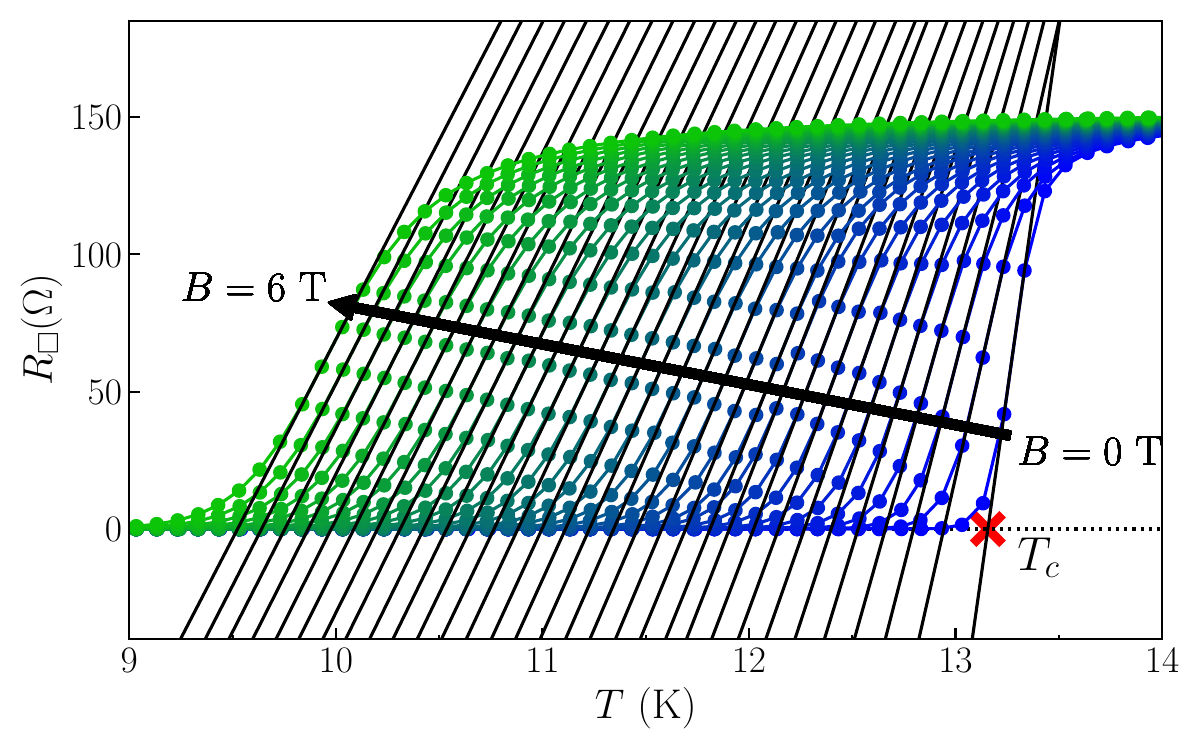}
	\caption{Magnetic field variation of the temperature-dependent sheet resistance $R_\square(T)$ for the 10 nm sample. Black lines are given by the maximal slope of $R_\square(T)$ curves and the temperature of the superconducting transition is determined by the intersect of the maximal slope line (black solid lines) and the zero resistance line (black dotted line).}
	\label{fig:low}
\end{figure}

\begin{figure}
	\centering
	\includegraphics[width=8.6cm]{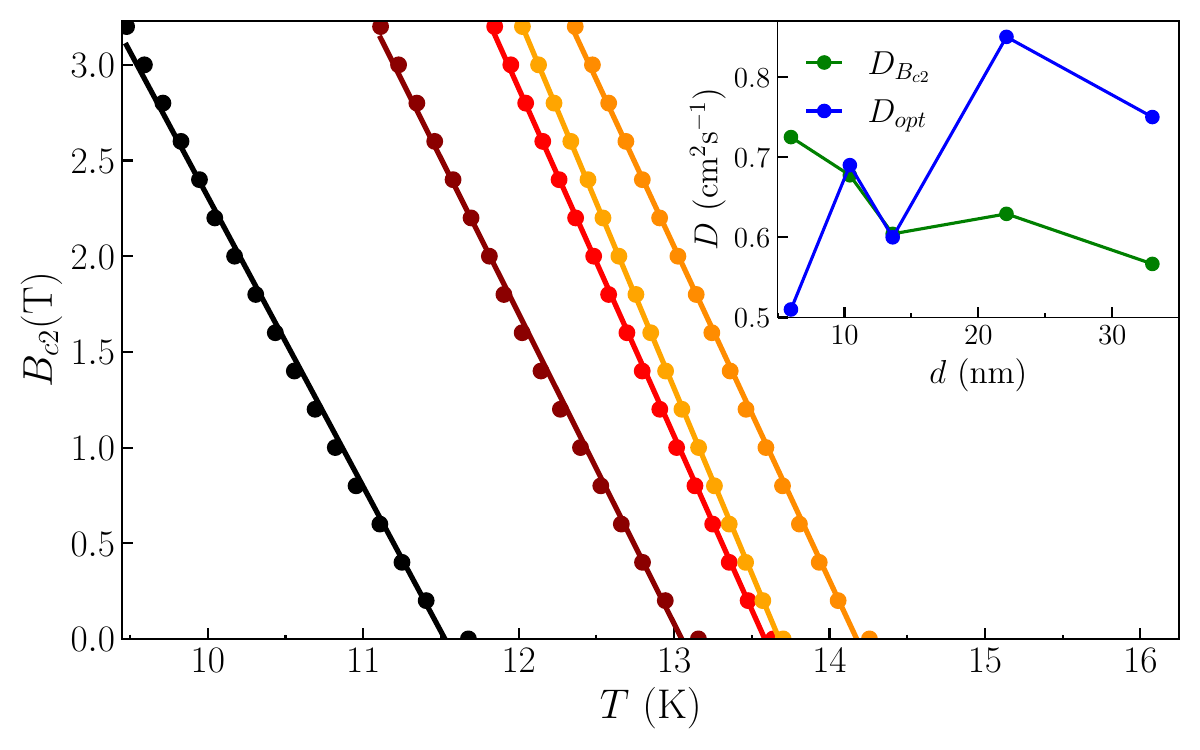}
	\caption{Temperature dependence of the upper critical field $B_{c2}$. The solid lines are linear fits to the $B_{c2}(T)$ data. The color-coding is same as in Fig.~\ref{fig:se}. The inset shows a comparison of the diffusivity obtained from the slope of $B_{c2}(T)$ (green) and that estimated from the proposed optical model (blue).}
	\label{fig:Tc}
\end{figure}

To verify the diffusivity estimated from the optical measurements, the magneto-resistance at low temperatures was measured (see Fig.~\ref{fig:low}),
and the temperature-dependence of the upper critical field $B_{c2}(T)$ was determined. We start with the Ginzburg-Landau (GL) result
for the upper critical magnetic field
\begin{equation}
   B_{c2} = \frac{\Phi_0}{2\pi\xi_{GL}^2(T)},
   \label{eq:GLBc}
\end{equation}
where $\Phi_0$ is the magnetic flux quantum and $\xi_{GL}$ is the GL coherence length. In the dirty limit, $\xi_{GL}(T)$ satisfies
\begin{equation}
   \frac{1}{\xi_{GL}^2(T)}=\frac{1}{0.855^2\xi_0 l}\frac{T_c-T}{T_c},
   \label{eq:XiGL}
\end{equation}
where $\xi_0 = \hbar v_F/(\pi\Delta)$ is the BCS coherence length, $T_c$ is the superconducting critical temperature, and $\Delta$ is the superconducting gap.
Recalling the BCS relation $\Delta=1.764\ k_BT_c$, one
can express the diffusivity via the temperature derivative of $B_{c2}$ as\cite{bartolf2015fluctuation}
\begin{equation}
   D_{B_{c2}} = -\frac{4k_B}{\pi e}\left(\frac{ \partial B_{c2}}{\partial T}\right)^{-1}_{T_c}.
   \label{eq:DBc}
\end{equation}
The diffusivities were estimated from the slope of  $B_{c2}(T)$ curves showed in Fig.~\ref{fig:Tc}. Comparison in the inset of Fig.~\ref{fig:Tc} shows that the diffusivity $D_{B_{c2}}$ estimated from the magneto-resistance is comparable to  $D_{opt}$, calculated from the optical model. The expected decrease
of the diffusivity at low thicknesses is present in  $D_{opt}$, but $D_{B_{C2}}$ is increasing with lowering the film thickness. This
paradoxical behaviour was likewise observed in Ref.~\citenum{ezaki2012localization}. In Ref.~\citenum{shoji1992superconducting} relation (\ref{eq:DBc}) was corrected to take
into account that NbN is supposed to be a strong coupling superconductor. However, this would lead to further
increase of $D_{B_{c2}}$. Alternatively, in Ref. \citenum{herman2016microscopic} the authors showed that the broadened tunneling spectra of dirty superconductors, also known as Dynes superconductors,  such as NbN\cite{noat2013unconventional,chaudhuri2013niobium}, can be explained
by the presence of two types of scattering processes, namely the pair-conserving and the pair-breaking scattering.  The rate of these scatterings is $\Gamma$ and $\Gamma_D$, respectively. They also calculated the thermodynamic properties
of these superconductor, expressing the GL coherence length as\cite{herman2018thermodynamic}
\begin{equation}
   \frac{1}{\xi_{GL}^2(T)}=\frac{12[1-\zeta(2,\frac{1}{2}+\alpha)]}{\pi\zeta(2,\frac{1}{2}+\alpha)}\frac{\hbar\Gamma}{\Delta}\frac{k_B(T_c-T)}{\Delta}\frac{1}{\xi_0^2},
   \label{eq:XiHH}
\end{equation}
where $\alpha=\hbar\Gamma_D/(2\pi k_BT_c)$ and $\zeta(s,x)$ is the Hurwitz zeta function. Then, diffusivity can be expressed as
\begin{equation}
D^\prime_{B_{c2}} = -K(\alpha)\frac{k_B}{e}\left(\frac{ \partial B_{c2}}{\partial T}\right)^{-1}_{T_c}\!\!\!,
\label{eq:DBcK}
\end{equation}
where $K=2\pi[1-\zeta(2,\frac{1}{2}+\alpha)]/\zeta(2,\frac{1}{2}+\alpha)$ and can be approximated by the
expression
\begin{equation}
K(\alpha) \approx \frac{1.273+1.155\alpha}{1+2.432\alpha+2.206\alpha^2}.
\label{eq:K}
\end{equation}
For $\alpha=0$, the function $K(\alpha)$ reduces to the BCS value $4/\pi$, however, for thin films, $\alpha$ is
non-zero and thus decreases the diffusivity. Comparing the diffusivity values $D_{B_{c2}}$ for the 6 nm sample and 30 nm samples, $\alpha = 0.15$ was estimated. This yields an estimate of $\hbar\Gamma_{\textrm{D}} = 0.9$ meV, which is reasonable for thin NbN films. $\hbar\Gamma_{\textrm{D}}$ estimated from tunneling spectra varies in the range
from negligibly small values up to tenths of meV, and even as large as $\Delta/2$ for strongly disordered samples with suppressed $T_c$.\cite{vsindler2014far,noat2013unconventional,henrich2014influence, carbillet2020spectroscopic,chand2012transport, kamlapure2013emergence}

The electron mean free
path was estimated as $l = \sqrt{ \hbar/ \mathcal{Q}\Gamma m_e}\approx2-3~$\AA\ is slightly above half of the lattice parameter $a=4.4~$\AA, which indicates that the samples are close to the Ioffe-Regel limit $k_Fl\to1$. The Ginzburg-Landau coherence length $\xi_{GL}(0)$ estimated by both Eqs. (\ref{eq:XiGL}) and (\ref{eq:XiHH}) ranges from $3~$nm to $4~$nm, which agrees with
a commonly measured value in thin NbN films.\cite{engel2006electric,jesudasan2011upper,shoji1992superconducting}

\section{Discussion}
\label{ch:disc}

It is now convenient to compare the revisited parameters of NbN with DFT simulations. We estimated the density of carriers in NbN as $n=\sigma_0\Gamma m_e/e^2= 9\times10^{28}~m^{-3}=2~V^{-1}_{\textrm{f.u.}}$, where $V_{\textrm{f.u.}}$ is the formula unit volume. For $\delta$-NbN, $V_{\textrm{f.u.}}=a^3/4$, where $a=4.4~\textrm{\AA}$ is the lattice constant. This result is in excellent agreement with the value obtained by integrating the DFT DOS from the threshold of the peak responsible for the modeled inter-band transition ($\approx E_F - 4~$eV) up to $E_F$.\cite{babu2019electron,amriou2003fp,sanjines2006electronic,papaconstantopoulos1985electronic,schwarz1977electronic,palanivel1993electronic} This also contradicts the common assumption, that all 4 Nb d-orbital electrons are conducting.\cite{chand2012transport} We should emphasize here, that the agreement of our estimation of the number of electrons in the conductive band with the DFT result
was obtained despite the fact that the DFT DOS itself is two times smaller than the "Drude" DOS, which we determined from the Einstein relation $N(E_F)=\sigma_0/(e^2D_{opt})$. This can be explained by the fact that the determined bottom of the conductive band $E_c=E_F-2~$eV is much closer to $E_F$ compared to DFT result $4~$eV. The value of $E_F-E_c$ was estimated from free electron relation between the density of electrons and DOS, given as $n = 2N(E_F)(E_F-E_c)/3$. Similar effect was observed in ARPES measurement where the conductive band was significantly flattened in comparison to the DFT calculation, naturally leading to higher DOS at the Fermi level.\cite{yu2021momentum} The $1~$eV scale of smearing of the ARPES
bandstructure also suggests a high value of scattering rate $\Gamma$.
Nevertheless, ARPES agrees with DFT on the value of the Fermi momentum
being approximately half of the $\Gamma$L path with length $\sqrt{3}\pi/a=1.24$~\AA$^{-1}$. This agrees with the estimated $k_F\approx 0.6$~\AA$^{-1}$.

For the "Drude" DOS $N(E_F)$, we obtained 2 states of both spins/eV/$V_{\textrm{f.u.}}$. Utilizing $\sigma_{DC}$ instead of $\sigma_0$, would lead to a different DOS, $N^\prime=(1-\mathcal{Q}^2)N(E_F)$ which is thickness-dependent.\cite{sidorova2020electron, chockalingam2008superconducting}
The authors of Ref.~\citenum{chockalingam2008superconducting} successfully compared this thickness-dependent $N^\prime$, obtained from transport measurements, to the DFT-calculated value, although the two are inherently different in nature.
Furthermore, it is important to emphasize that DFT calculations only partially account for electron-electron and electron-phonon interactions, both of which are significant in NbN.\cite{geballe1966high} Therefore, comparing the DFT value of $0.8–1.2$ $\textrm{eV}^{-1}\textrm{V}_{\textrm{f.u.}}^{-1}$ with $N^\prime=0.5-1~\textrm{eV}^{-1}\textrm{V}_{\textrm{f.u.}}^{-1}$, or even with $N(E_F)\approx2~\textrm{eV}^{-1}\textrm{V}_{\textrm{f.u.}}^{-1}$, is questionable. Nevertheless, $N(E_F)$ and
$N^\prime$ can be considered upper and lower bound, respectively, for the actual DOS.

\section{Conclusion}
\label{ch:con}

In conclusion, we argue that the quantum corrections to the conductivity of NbN films are present at optical frequencies and significantly alter their dielectric function. Therefore, we analyze their optical conductivities, utilizing the quantum-corrected Drude-Lorentz model. The proposed model yields to an excellent fit to the $\sigma(\omega)$ and provides parameters of the electronic fluid such as: the electron concentration $n$, the diffusion coefficient $D$, the Ioffe-Regel parameter $k_Fl$, and the electronic density of states $N(E_F)$. The obtained diffusion coefficient agrees with the magneto-transport measurement, moreover, the estimated $n$ is consistent with ab initio simulations. The determined electron relaxation rate $\hbar\Gamma\approx 1.8~$eV, consistent with the presence of high disorder in NbN films, is an order of magnitude higher than the commonly considered value obtained from standard optical models. This emphasizes the importance of quantum corrections in the analysis. Moreover, various puzzling phenomena like the double ENZ\cite{ran2021stoichiometry}, the increase of sheet resistance at lower thicknesses, and inconsistencies in electron relaxation rates are explained by this model. For other reported effects, such as increasing diffusivity with lowering of the thickness, and high electron density of states, we have suggested explanations, which could be verified by further experiments.
\subsection{Acknowledgments}
We thank to R. Hlubina and R. Marto\v{n}\'{a}k for useful discussions. 
This work was supported by the Slovak Research and Development Agency under the contracts APVV-20-0425, by the QuantERA grant SiUCs and by the project skQCI (101091548), founded by the European Union (DIGITAL) and the Recovery and Resilience Plan of the Slovak Republic. M. Pol\'{a}\v{c}kov\'{a} was supported by the Slovak Research and Development Agency under Contract No. APVV-19-0303.

It is also the result of support under the Operational Program Integrated Infrastructure for the projects: Advancing University Capacity and Competence in Research, Development and Innovation (ACCORD, ITMS2014+:313021X329) and UpScale of Comenius University Capacities and Competence in Research, Development and Innovation (USCCCORD, ITMS 2014+:313021BUZ3), co-financed by the European Regional Development Fund.

\appendix

\section{Mid-infrared range optical properties and samples aging}
\label{app:mir}
The mid-IR measurements presented in the main text were performed approximately one year after the original set of measurements. During this period, the samples experienced slight degradation, as indicated by increased sheet resistance (see Tab. \ref{tab:deg}). Consequently, we repeated the visible range ellipsometry on the samples, providing the conductivities shown in Fig. \ref{fig:se} b). The visble-range data were fitted to the proposed model (thin lines), with the resulting parameters listed in Table \ref{tab:params2}. The extrapolations from the newly acquired visible range data remain largely consistent with the results from the original analysis, and the extrapolated values also match the mid-IR data.

The degradation of the samples manifested as an increase in sheet resistance, with the most significant changes occurring in the thicker samples. The thinner samples exhibited minimal degradation, allowing for a more direct comparison with the original data. The parameter that showed the most notable change is 
	$\mathcal{Q}$, which increased systematically across all samples. Again, this change was more significant for the thicker samples.

\begin{table}
	\centering
	\renewcommand{\arraystretch}{1.6}
	\begin{tabular}{>{\centering\arraybackslash}p{1.9cm} ||>{\centering\arraybackslash}p{1.0cm} >{\centering\arraybackslash}p{1.3cm} >{\centering\arraybackslash}p{0.9cm} >{\centering\arraybackslash}p{1.1cm} >{\centering\arraybackslash}p{1.1cm}
		}
		\hline 
		$d~(nm)$& 6.0 & 10.4 &  13.6 & 22.1  & 33.0 \\ 
		\hline
		$R_\square^{1\textrm{yr}}~(\Omega)$  & 350  & 143 & 118 & 61 & 48 \\ 
		$\Delta R_\square/R_\square~(\%)$ & 3 & 5 & 10 & 5 & 17\\
		
		\hline
	\end{tabular}
	\caption{Change in the sheet resistance of thin NbN films after passing 1 year.}
	\label{tab:deg}
\end{table}

\begin{table}[h!]
	\centering
	\renewcommand{\arraystretch}{1.5}
	\begin{tabular}{>{\centering\arraybackslash}p{0.8cm} |>{\centering\arraybackslash}p{1.5cm} >{\centering\arraybackslash}p{1.0cm} >{\centering\arraybackslash}p{0.7cm} >{\centering\arraybackslash}p{1.5cm} >{\centering\arraybackslash}p{1.1cm}
			>{\centering\arraybackslash}p{1.0cm} }
		\hline
		$d$& $\sigma_0  $ & $\hbar\Gamma$ &  $\mathcal{Q}$& $\sigma_1 $ & $\hbar\Gamma_1$& $\hbar\omega_1$\\ 
		
		$(\textrm{nm})$& $(S\mu\textrm{m}^{-1})$ & $(\textrm{eV})$ &  $(1)$& $(S\mu\textrm{m}^{-1})$ & $(\textrm{eV})$& $(\textrm{eV})$\\ 
		\hline\hline
		6.0  & 0.88 & 1.83 & 0.78 & 0.63 & 2.44 & 5.28\\ 
		10.4 & 0.94 & 1.79 & 0.59 & 0.77 & 2.02 & 5.40\\
		13.6 & 1.01 & 1.70 & 0.70 & 0.87 & 1.24 & 4.93\\
		22.1 & 0.95 & 1.84 & 0.64 & 0.72 & 1.88 & 5.24\\
		33.0 & 0.95 & 1.73 & 0.66 & 0.66 & 3.08 & 6.55\\
		\hline
	\end{tabular}
	\caption{Parameters of optical model (\ref{eq:TheModel}) and (\ref{eq:im}) resulting from fitting procedure applied to visible range SE repeated after 1 year.}
	\label{tab:params2}
\end{table}

\section{Extraction of optical constants from ellipsometry}
\label{app:eoc}
In spectroscopic ellipsometry, the measured quantity is the ratio $\rho$
of the complex reflection coefficients of the p- and the s-polarized light, typically expressed via angles $\Psi$ and $\Delta$\cite{fujiwara2007spectroscopic}
\begin{equation}
\rho\equiv\tan(\Psi)e^{i\Delta}\equiv\frac{r_p}{r_s}.
\label{eq:rho}
\end{equation}
The system under study is modelled as a semi-infinite sapphire substrate, with refractive index $n_s(\omega)$, on top of which is a thin film with refractive index $n_f(\omega)$ and vacuum above it.  
The reflection coefficients of the vacuum-film-substrate system can be expressed\cite{fujiwara2007spectroscopic}
\begin{equation}
\begin{aligned}
r_s = \frac{r_{vf,s}+r_{fs,s}e^{-2i\beta}}{1+r_{vf,s}r_{fs,s}e^{-2i\beta}},\\
r_p = \frac{r_{vf,p}+r_{fs,p}e^{-2i\beta}}{1+r_{vf,p}r_{fs,p}e^{-2i\beta}}. 
\label{eq:r_s,p}
\end{aligned}
\end{equation}
Here, the phase difference across the thin film is
\begin{equation}
\beta=\omega d n_f \cos(\theta_1)/c,
\end{equation}
where $d$ is the thin film thickness, $\theta_1$ is the angle of refraction which can be obtained from the Snell's law $\sin(\theta_1)=\sin(\theta_0)/n_f$ and the angle of incidence $\theta_0$. The reflection coefficient for the s-/p-polarized light on the vacuum-film and film-substrate interfaces are denoted as $r_{vf,s/p}$ and $r_{fs,s/p}$, respectively.
Utilizing Fresnel equations, these coefficients can be expressed
\begin{equation}
\begin{aligned}
r_{vf,s}=\frac{\cos(\theta_0)-n_f\cos(\theta_1)}{\cos(\theta_0)+n_f\cos(\theta_1)},\\
r_{vf,p}=\frac{n_f\cos(\theta_0)-\cos(\theta_1)}{n_f\cos(\theta_0)+\cos(\theta_1)},\\
r_{fs,s}=\frac{n_f\cos(\theta_1)-n_s\cos(\theta_2)}{n_f\cos(\theta_1)+n_s\cos(\theta_2)},\\
r_{fs,p}=\frac{n_s\cos(\theta_2)-n_f\cos(\theta_1)}{n_s\cos(\theta_1)+n_f\cos(\theta_2)}.
\label{eq:rs}
\end{aligned}
\end{equation}
Here, the second angle of refraction $\theta_2$ can be obtained again from the Snell's law. This way, the 
complex ratio
\begin{equation}
\rho=\rho(d,\theta_0,n_f,n_s)
\label{eq:rho}
\end{equation}
 is a function of the film thickness $d$ and the refractive indices $n_f$, $n_s$. Thus, for known film thickness $d$ and known index $n_s$, the refractive index $n_f$ of the thin films can be obtained by numerical inversion of (\ref{eq:rho}) with respect to $n_f$.  

\section{Effective medium theory}
\label{app:emt}

In Ref.~\citenum{bower2021tunable}, the double ENZ was explained via Maxwell-Garnet (MG) effective medium theory,
which describes the optical properties of composite materials consisting of polarizable inclusions in an insulating
matrix. In the case of NbN, NbN nanoparticles are immersed in a matrix of insulating niobium oxides. Taking the dielectric function of the oxide from Ref.~\citenum{o2014electronic}, the resulting MG
formula provides $\epsilon(\omega)$ with two zeros. However, such effective medium with an insulating matrix is inevitably an insulator. Moreover, the presence of oxygen in the NbN films was associated with the degree of porosity
and it manifests itself in a significant increase of the resistance and/or the residual-resistance ration (RRR).\cite{cabanel1990correlations} Considering the sheet resistance and RRR (see Table~\ref{tab:meas}), our samples have negligible oxygen content. The presence of oxygen should affect the density of the films as well,\cite{cabanel1990correlations} which was estimated from  XRR, to have a constant value of $7.8~$gcm$-3$, which is close the ideal cubic NbN value.  

If the matrix would be a bad conductor, with a conductivity peak at 5-7 eV, the MG model
indeed reproduces both the suppression of the DC conductivity and the double ENZ behaviour. But, as can be seen in Fig.~\ref{fig:ds}a, by varying the volume fraction of the inclusion, the model interpolates between the Drude conductivity and the conductivity of the poorly conducting matrix, i.e., the
conductivity varies in the whole frequency range. Besides the DC conductivity suppression, $\sigma(\omega)$ would vary at energy $\approx\hbar\Gamma$, too, and the optical peak would change rapidly its weight. None of this
behaviour was observed in the SE measurements.

\section{Drude-Smith model}
\label{app:DSM}
\begin{figure}
	\centering
	\includegraphics[width=8.6cm]{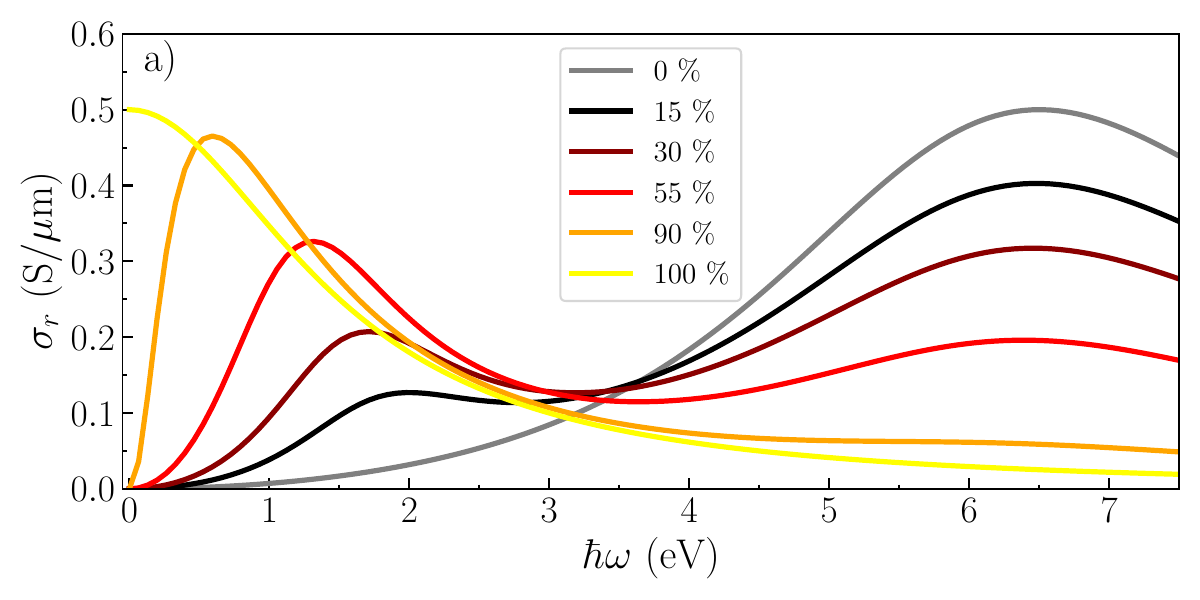}
	\includegraphics[width=8.6cm]{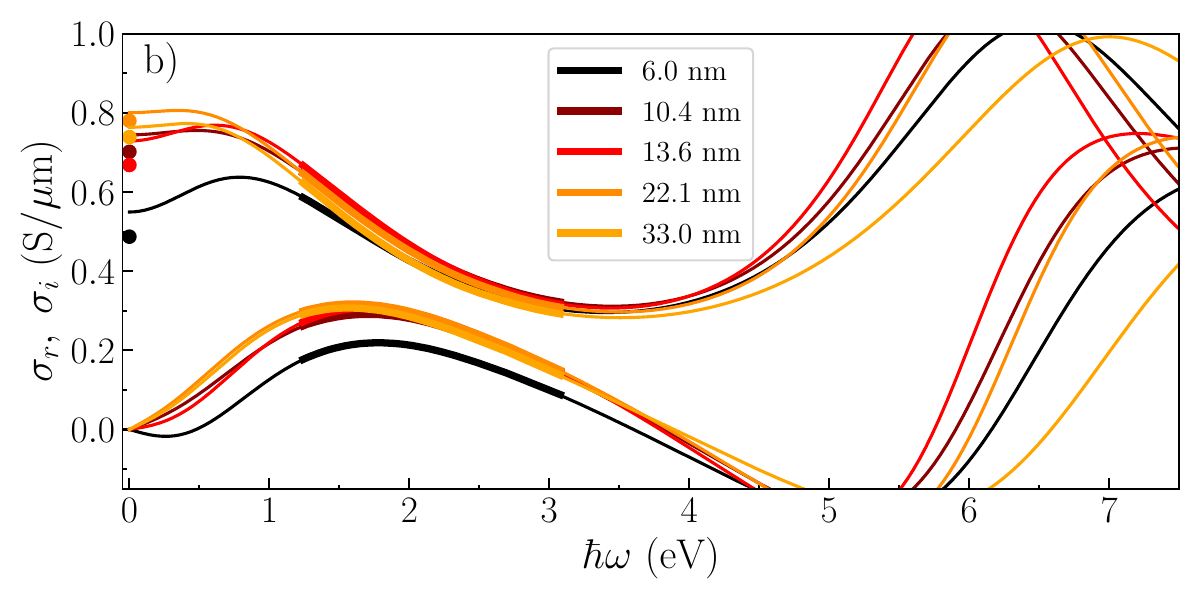}
	\caption{a) Prediction of Maxwell-Garnet theory for a metallic inclusion in a niobium oxide matrix for various volume fractions of the inclusion. b) Thin lines are the Drude-Smith model curves obtained as best fit to data from ellipsometry depicted by thick lines. Points at zero frequency are the measured DC conductivities.}
	\label{fig:ds}
	
\end{figure}
Another approach which yields to the anomalous Drude peak is the Drude-Smith model, which is based
on a material formed by granules whose boundaries cause reflection of electrons. This model, as derived in Ref.~\citenum{cocker2017microscopic},
leads to the following corrections to the Drude formula 
\begin{equation}
\sigma_{DS}(\omega) = \frac{\sigma_{0} }{1-i\omega/\Gamma_{\textrm{diff}} }\left(1-\frac{c}{1-i\omega/a}\right).
\label{eq:DS}
\end{equation}
Here, the parameters $\Gamma_{\textrm{diff}}$ and $a$ are determined by the granule size $L$, utilizing the following relations
\begin{equation}
\Gamma_{\textrm{diff}}=\frac{1}{\tau}+\frac{2v_{th}}{L}
,\quad
a=\frac{12v_{th}}{L}\frac{\tau}{L/v_{th}+2\tau},
\label{eq:a}
\end{equation}
where $v_{th}=\sqrt{k_BT/m_e}$ is the thermal velocity. We added the inter-band transition term to this model and the high frequency permittivity $\epsilon_\infty=1.62$ contribution in the imaginary part is considered, too.  resulting in
\begin{equation}
\begin{aligned}
\sigma(\omega) =  \frac{\sigma_{0} }{1-i\omega/\Gamma_{\textrm{diff}} }\left(1-\frac{c}{1-i\omega/a}\right) +\\ \frac{\sigma_1}{1+i\frac{\omega_1^2-\omega^2}{\Gamma_1\omega}} -i(\epsilon_\infty-1)\omega\epsilon_0.
\label{eq:DS_full}
\end{aligned}
\end{equation}
This model fits our experimental data very well (see Fig.~\ref{fig:ds}b). However, the resulting granule size is $5~$\AA, which is at least one order of magnitude smaller than the grain size estimated by X-Ray diffraction (XRD) measurement. For reasonable parameters (the measured grain size $L\approx10~$nm and relaxation rate $1/\tau=\Gamma\approx2~\textrm{eV}/\hbar$),
this model leads to a displaced Drude peak in THz frequency range (meV). Thus, the fit of this model to our data produces unreasonable parameters. We do not claim
that this effect is not present as it still can play a role at much smaller energies, where our measurements are not sensitive. 
\begin{table}
	\centering
	\renewcommand{\arraystretch}{1.5}
	\begin{tabular}{>{\centering\arraybackslash}p{0.75cm} |>{\centering\arraybackslash}p{1.3cm} >{\centering\arraybackslash}p{0.9cm} >{\centering\arraybackslash}p{0.8cm} >{\centering\arraybackslash}p{0.8cm}>{\centering\arraybackslash}p{1.3cm} >{\centering\arraybackslash}p{0.8cm}
			>{\centering\arraybackslash}p{0.75cm} }
		\hline
		$d$& $\sigma_{0}  $ & $\hbar\Gamma_{\textrm{diff}}$ &  $c$&$\hbar a$& $\sigma_1$ & $\hbar\Gamma_1$& $\hbar\omega_1$\\ 
		
		$(\textrm{nm})$& $(S\mu\textrm{m}^{-1})$ & $(\textrm{eV})$ &  $(1)$&$(\textrm{eV})$& $(S\mu\textrm{m}^{-1})$ & $(\textrm{eV})$& $(\textrm{eV})$\\ 
		
		\hline\hline
		6.0  & 1.00 & 1.38 & 0.48 & 1.08 & 1.00 & 0.76 & 5.52\\ 
		10.4 & 1.17 & 1.36 & 0.38 & 1.31 & 1.01 & 0.72 & 5.50\\
		13.6 & 1.18 & 1.35 & 0.40 & 1.11 & 1.08 & 0.68 & 5.50\\
		22.1 & 1.17 & 1.32 & 0.34 & 1.36 & 1.05 & 0.76 & 5.70\\
		33.0 & 1.17 & 1.24 & 0.37 & 1.20 & 0.97 & 0.91 & 6.00\\
		\hline
	\end{tabular}
	\caption{Paremeters of Drude-Smith model (\ref{eq:DS}) providing best fit to the experimental data.}
	\label{tab:paramsDS}
\end{table}
\section{Approximative formula for imaginary part of the modelled conductivity}
\label{app:app}
\begin{figure}
	\centering
	\includegraphics[width=8.6cm]{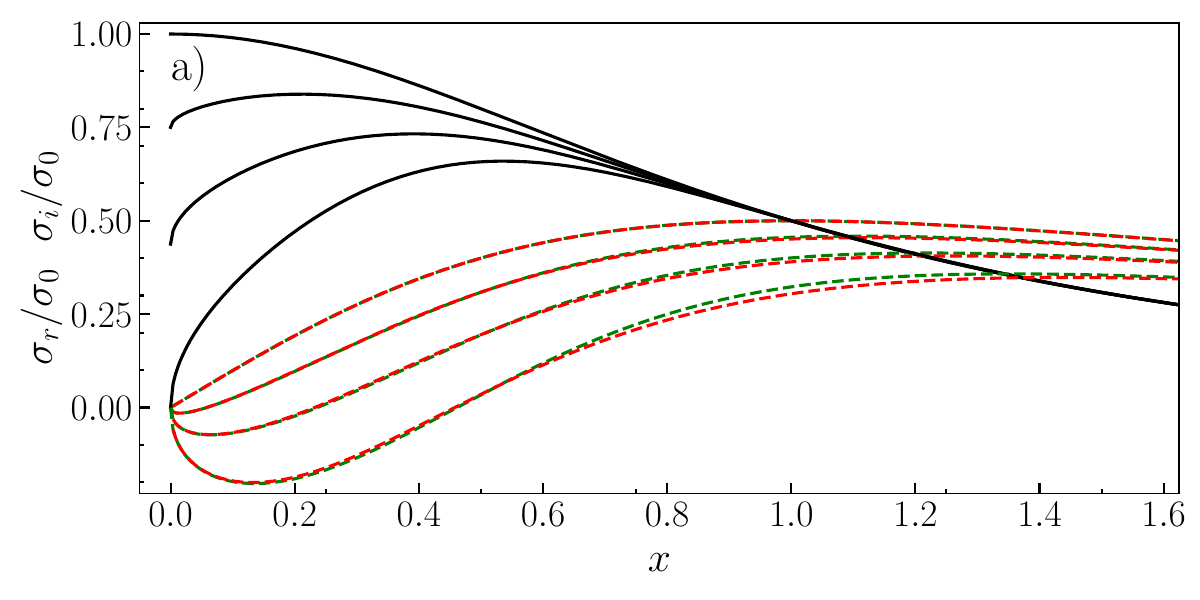}
	\includegraphics[width=8.6cm]{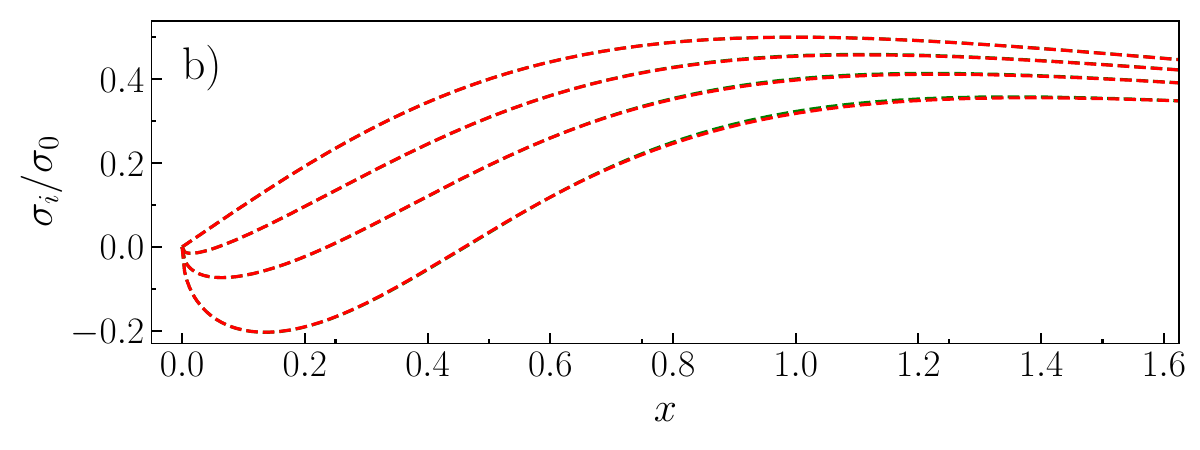}
	\caption{a) Black lines are the modified Drude conductivity $\sigma_r(x)$ with quantum corrections of various strength $\mathcal{Q}=0,\ 0.5,\ 0.75,\ 1$. Green dashed lines are their corresponding Kramers-Kronig images obtained numerically. Red dashed lines are given by the formula (\ref{eq:imx1}) for the Hilbert transforms of $\tilde{\sigma}_r(x)$. b) Green dashed lines are the same as in a), red dashed lines are plots of the formula (\ref{eq:imx2}) for $\mathcal{H}[\tilde{\tilde\sigma}_r(x)]$.}
	\label{fig:hilli}
\end{figure}

The Drude-Lorentz and Drude-Smith models are convenient because both the
real and the imaginary part are accessible in a simple closed formula. Therefore, they can be easily
implemented in a fitting procedure, which are computationally less demanding. More complex models,
typically expressing one part of the dielectric function, usually require to compute the other one
numerically, as we have indeed done for the proposed model (\ref{eq:TheModel}). However, we derived a simple approximative analytical formula for the
KK image of $\sigma_{\textrm{QCD}}(\omega)$, too. We start by introducing the dimensionless frequency
$x=\omega/\Gamma$ and expanding the exponential function
\begin{equation}
e^{-2x^2} = \Big(1+2x^2+\frac{1}{2}(2x^2)^2+...\Big)^{-1}.
\label{eq:expansion}
\end{equation}
Taking the first two terms, we obtained the approximative form of $\sigma_{\textrm{QCD}}(x)$
\begin{equation}
\tilde{\sigma}_{\textrm{QCD}}(x) = \frac{\sigma_0}{1+x^2}\Big(1-\mathcal{Q}^2\frac{1-\sqrt{x}}{1+2x^2}\Big).
\label{eq:appsigma}
\end{equation}
Utilizing the fact that the real part of the conductivity is an even function of $x$, the Hilbert transform of Eq.~(\ref{eq:appsigma}) is
\begin{equation}
\mathcal{H}[\tilde\sigma_{\textrm{QCD}}(x)] = \frac{2x}{\pi}\mathrm{P}.\mathrm{V}.\int_0^\infty\frac{\tilde\sigma_{\textrm{QCD}}(s)}{x^2-s^2}dx.
\label{eq:H}
\end{equation}
For which we obtained
\begin{equation}
\mathcal{H}[\tilde\sigma_{\textrm{QCD}}(x)] = \frac{\sigma_0 x}{1+x^2}\Big(1-\mathcal{Q}^2\frac{ax^2-b+1/\sqrt{x}}{1+2x^2}\Big),
\label{eq:imx1}
\end{equation}
where
\begin{equation}
\begin{aligned}
a &= 2(2\sqrt{2}-2^{3/4}-1)\approx0.293,\\
b& = (1-3\sqrt{2}+2^{7/4})\approx0.121.
\end{aligned}
\label{eq:imx1ab}
\end{equation}
The comparison of Eq.~(\ref{eq:imx1}) (red dashed lines) with the numerical result (green dashed lines) for various values of quantumness $\mathcal{Q}=0,\ 0.5,\ 0.75,\ 1$ is in Fig.~\ref{fig:hilli}a. One can see a excellent match, except for the value $\mathcal{Q} = 1$, where a slight disagreement can be seen. This can be treated by taking the next term in the expansion (\ref{eq:expansion}). We calculated the Hilbert transform of the function
\begin{equation}
\tilde{\tilde\sigma}_{\textrm{QCD}}(x) = \frac{\sigma_0}{1+x^2}\Big(1-\mathcal{Q}^2\frac{1-\sqrt{x}}{1+2x^2+2x^4}\Big),
\label{eq:imx2}
\end{equation}
with the result
\begin{equation}
\begin{aligned}
\mathcal{H}[\tilde{\tilde\sigma}_{\textrm{QCD}}(x)] &=\frac{\sigma_0 x}{1+x^2}\Big(1+\mathcal{Q}^2\Big(\sqrt{2}-\\
&\frac{(\alpha x^2+\beta)(x^2+1)+1+1/\sqrt{x}}{1+2x^2+2x^4}\Big)\Big),
\end{aligned}
\label{eq:imx2ab}
\end{equation}
where
\begin{equation}
\begin{aligned}
\alpha = &2-2^{7/4}\sin\Big(\frac{\pi}{8}\Big)+\frac{8}{\pi}\int_0^\infty\!\!\!\frac{x^{5/2}dx}{1+2x^2+2x^4}\approx 3.136,\\
\beta & = 2^{5/4}\sin\Big(\frac{\pi}{8}\Big)- \frac{4}{\pi}\int_0^\infty\!\!\!\frac{x^{1/2}dx}{1+2x^2+2x^4}\approx 0.308.\\
\end{aligned}
\label{eq:imx}
\end{equation}
In Fig.~\ref{fig:hilli}b we compare numerical transformations of Eq.~(\ref{eq:TheModel}) (green dashed lines) to the approximative formula (\ref{eq:imx2}) (red dashed lines), indicating deviations smaller than $1~\%$.  
Finally, fitting the experimental data with the approximate formula for the imaginary part Eq.~(\ref{eq:imx}) produces identical results as the numerical transformation does.


\section{Electron structure and $\epsilon_\infty$ estimation}
\label{app:elec}
The picture of bonding and the configuration for 10 niobium and nitrogen valence electrons in NbN were proposed by many authors, \cite{geballe1966high, schwarz1977electronic, schwarz1987band} and it was later largely confirmed by
the partial DOS from DFT calculations and X-ray photoelectron spectroscopy.\cite{babu2019electron, sanjines2006electronic} Namely, there is a complicated Fermi surface created by Nb's 4d orbitals occupied by two electrons.\cite{babu2019electron, pfluger1985dielectric}. This agrees with the
optical estimation of the Drude weight mentioned in the main text. Next, there is strong hybridization of Nb 4d and N 2p orbitals containing approx. 6 electrons forming a peak in DOS, approx. 6 eV below $E_F$, providing electrons for the modelled inter-band transition. A naive approach utilizing the strength and the width of the optical peak $Z = \sigma_1\Gamma_1m_e/(n_{NbN}e^2)$ gives 3-5 electrons.
Here, $n_{NbN}=1/V_{\textrm{f.u.}}$ is the concentration of formula units. The estimated number of electrons matches well with the band structure value, considering neglected joint DOS influence. Finally, the calculations indicate that
the remaining 2 electrons occupy bands low in energy (10-20 eV below $E_F$), corresponding to N 2s orbitals. Contribution to the optical response due to the transition of these electrons to Fermi level is included via the parameter $\epsilon_\infty$ together with transitions of relevant remaining core electrons.

Similarly to the estimation of the electron number in the DOS peak from the weight of the inter-band transition peak, we estimated the contribution $\epsilon_\infty$ to the dielectric function from high-energy transitions.
In Ref.~\citenum{neilinger2019observation}, the $\epsilon_\infty$ was expressed via the number of core electrons in their respective atomic level $k$, i.e. $Z_k$, as follows
\begin{equation}
\epsilon_\infty \approx 1+\sum_k \frac{Z_k\Omega_k^2}{\omega_k^2}, \quad \Omega_k^2 = \frac{n_{\scriptscriptstyle{NbN}}e^2}{m_e\epsilon_0}.
\label{eq:eps_inf}
\end{equation}
Here $\hbar\omega_k$ is the energy of the atomic level $k$ with respect to the Fermi level. The values of $\hbar\omega_k$ are listed in Ref.~\citenum{energies}. The relevant orbitals which are not too low in energy for niobium are 3s$^2$, 3p$^6$, 3d$^{10}$, and 4s$^2$, with energies 467 eV, 370 eV, 203 eV, and 56 eV, respectively. Nitrogen contributes with 1s$^2$ and 2s$^2$, with energies 410 eV and 37 eV, respectively. The energy of the nitrogen's 2s orbital is not taken from Ref.~\citenum{energies}, but instead, the value 15 eV was taken, which is suggested by the predictions on the NbN electronic band structure, summarized in the previous paragraph. Finally, we obtained $\epsilon_\infty = 1.62$.

\section{Sample preparation}
\label{ch:sampprep}

The thin NbN films were prepared by pulsed laser deposition (PLD, Omicron system with Coherent Compex Pro 201 F laser) by means of  a KrF laser with wavelength of 248 nm and pulse duration of 35 ns. The films were grown on c-cut sapphire substrates cleaned in ultrasonic bath in acetone, isopropyl alcohol, and deionized water in succession. The deposition was performed in high-vacuum chamber with the residual atmosphere pressure of $10^{-7}$ Pa. The ablation was carried out from a niobium target in N$_2+1\%$H$_2$ reactive atmosphere. The pressure of the atmosphere was 9.3 Pa and the substrate was heated up to 600°C.
For more details see Ref.~\citenum{volkov2019superconducting}.
\vspace{4cm}


\bibliographystyle{apsrev}
%

\end{document}